\documentclass[10pt,reprint,aps,prb,twocolumn,groupedaddress,showpacs]{revtex4-1} 
\usepackage{graphicx}
\usepackage{dcolumn}
\usepackage{bm}
\begin{document}
\newcommand{\hesan}{$^3$He }
\newcommand{\hesanend}{$^3$He}
\newcommand{\heyon}{$^4$He }
\newcommand{\heyonend}{$^4$He}
\newcommand{\kake}{$\times$}

\title{Effects of \hesan Impurity on Solid \heyon Studied by Compound Torsional Oscillator
}
\author{P. Gumann}
\altaffiliation[Present address: ]{Institute for Quantum Computing, University of Waterloo, ON, Canada.}\author{M.C. Keiderling} \author{D. Ruffner} \author{H. Kojima}\affiliation{Serin Physics Laboratory, Rutgers University,
Piscataway, NJ 08854 USA}
%
%
\begin{abstract}
Frequency shifts and dissipations of a compound torsional oscillator induced by solid \heyon samples containing \hesan impurity concentrations ($x_3$ = 0.3, 3, 6, 12 and 25 in units of 10$^{-6}$) have been measured at two resonant mode frequencies ($f_1$ = 493 and $f_2$ = 1164 Hz) at temperatures ($T$) between 0.02 and 1.1 K. The fractional frequency shifts of the $f_1$ mode were much smaller than those of the $f_2$ mode.  The observed frequency shifts continued to decrease as $T$ was increased above 0.3 K, and the conventional non-classical rotation inertia fraction was not well defined in all samples with $x_3 \geq$ 3 ppm.  Temperatures where peaks in dissipation of the $f_2$ mode occurred were higher than those of the $f_1$ mode in all samples.  The peak dissipation magnitudes of the $f_1$ mode was greater than those of the $f_2$ mode in all samples.  The activation energy and the characteristic time ($\tau_0$) were extracted for each sample from an Arrhenius plot between mode frequencies and inverse peak temperatures.  The average activation energy among all samples was 430 mK, and $\tau_0$ ranged from 2$\times 10^{-7}$ s to 5$\times 10^{-5}$ s in samples with $x_3$ = 0.3 to 25 ppm.  The characteristic time increased in proportion to $x_3^{2/3}$.  Observed temperature dependence of dissipation were consistent with those expected from a simple Debye relaxation model \emph{if} the dissipation peak magnitude was separately adjusted for each mode. Observed frequency shifts were greater than those expected from the model.  The discrepancies between the observed and the model frequency shifts increased at the higher frequency mode.
\end{abstract}
%
\maketitle
\section{introduction}
The discovery\cite{Kim04a,*Kim04b} that the resonant frequencies
of torsional oscillators (TOs) containing solid \heyon samples
increased at temperatures below about 300 mK has stirred much
excitement (see reviews\cite{Prokofev07,Balibar08,Galli08}).
Interpreted as a partial decoupling or so-called\cite{Leggett70}
``non-classical rotational inertia (NCRI)'' of the solid \heyon
samples, the discovery gave evidence for the
long-sought\cite{Meisel92} supersolid state of solid $^4$He
possessing simultaneously crystallinity and superfluid
properties.  In the temperature range where the observed TO
frequency shifts vary most rapidly, peaks in the energy
dissipations of the TO are also observed.  The frequency shifts
and the accompanying dissipations of TO have been confirmed in
many laboratories.\cite{Rittner06,Kondo07,Penzev07,Aoki07,Hunt09}
However, the interpretation in terms of supersolid state in the
loaded solid \heyon samples remains controversial and a
comprehensive understanding of the TO frequency shifts and other
related phenomena in quantum solid \heyon at low temperatures has
not been established.

There are puzzling observations in experiments designed to test
the supersolid interpretation.  Superflows that would be expected
to occur through supersolid \heyon samples under applied pressure
gradients have not been observed in the earlier flow
experiments\cite{Greywall77,Day07,Rittner09} carried out in the
same temperature range as the NCRI effect is observed.  Recently,
however, unusual mass flows\cite{Ray08,Ray10} were induced
through solid \heyon samples below about 0.6 K by applying
chemical potential gradients across the samples.  Propagation of
fourth sound that would be expected\cite{Andreev69,Sears10} for a
superfluid has not been detected.\cite{Aoki08b,Kwon10}  The shear
modulus of a thin solid \heyon slab was found\cite{Day07} to
"stiffen" in a very similar manner as the frequency of TO
containing an annular solid \heyon sample was found\cite{Kim04b}
to increase with decreasing temperature.  This similarity
indicated a common origin of these two effects.  While almost
none of these observations are clearly understood, it has become
apparent that the details of the observations are affected
strongly by the solid \heyon sample quality depending on growth
condition, sample geometry and size, and \hesan impurity.  The
objective of the present work is to gain understanding of the
puzzling role played by \hesan impurity by use of our compound TO
techniques.

The surprisingly high sensitivity of observed frequency shifts to
minute \hesan impurity concentrations ($x_3$) at parts per
million (ppm) levels in the \heyon samples has been reported by
Kim, et al. \cite{Kim04a,Kim08}  The temperature dependence of
the shear modulus was also found\cite{Day07} to depend on $x_3$.
On the other hand, a thermodynamic anomaly in heat capacity
occurs at temperatures which are relatively insensitive to the
value of  $x_3$.  The dynamics of \hesan atoms within solid
\heyon may be probed by NMR experiments. Simultaneous
observations of NMR and TO effects, of the same crystal, have
been carried out by Toda, et al.\cite{Toda10, Toda11} in solid
\heyon samples with $x_3$ down to 10 ppm.  They found three
different spin-lattice relaxation times suggesting the existence
of three different states of \hesan atoms in the solid matrix of
\heyonend.  Kim, et al.\cite{Kim_SS10a,Kim_SS10b} reported their
measurements of NMR relaxation times, $T_1$ and $T_2$, of \hesan
contained in solid \heyon samples with $x_3$ down to 16 ppm.  The
relation between all of these NMR observations and the role of
\hesan impurity in the supersolid phenomenon is yet to be
clarified.

One goal of our work is to measure how the dissipations
accompanying frequency shifts of solid-$^4$He-loaded TO depend on
the amount of added \hesan impurity.  The earlier
reports\cite{Kim04a,Kim08} focused mainly on the variation of
frequency shifts as $x_3$ was changed.  Another goal is to
measure how the dependence of frequency shifts and dissipations
on the TO frequency varies as $x_3$ is changed.  Frequency
dependence studies are difficult in single resonance mode TOs.
We report on the first systematic measurements of the dependence
of dissipations and frequency shifts on $x_3$ (0.3 $\sim$ 25 ppm)
using our compound TO having two resonance mode frequencies
($\sim$500 and $\sim$1200 Hz).  Unlike in the earlier \hesan
impurity dependence study \cite{Kim08}, all samples are grown in
the same torsional oscillator following nearly identical solid
growth procedures such that variance due to sample cell geometry
and sample quality would be minimized.  Measurements are made
simultaneously at the two frequencies under identical sample
conditions and they allow us to explore dynamical effects which
cannot be probed in a single mode TO.  For a given sample, the
maximum in dissipation of the higher frequency mode appears at a
higher (peak) temperature than the lower frequency mode.  As
$x_3$ is increased, the peak temperatures for both modes
increase.  Arrhenius plots of the mode frequencies vs. inverse
peak temperatures are analyzed to extract the activation energy
and the characteristic time involved in the dissipation process.
A simple form of Debye dissipation combined with the extracted
characteristic time is inadequate to represent the observed
frequency dependent dissipations of the two modes separately.
The frequency shifts expected from the dissipation by the
Kramers-Kronig relation cannot account for those observed in the
two modes.  The limiting characteristic time $\tau_0$ extracted
from the two modes varies in proportion to $x_3^{2/3}$.

\section{experiment}
\begin{figure}
\includegraphics[width=3in]{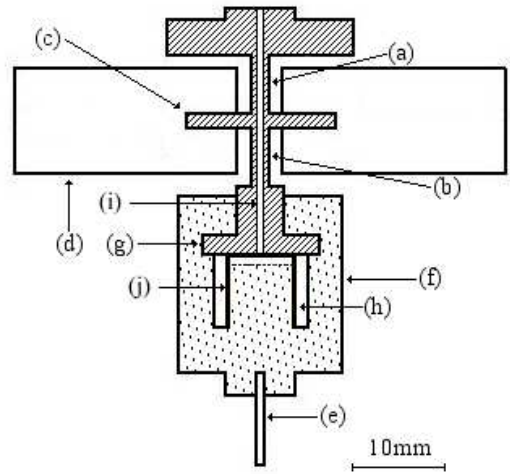}\\
\caption{Schematic of compound torsional oscillator.  (a) upper
torsion rod, (b) lower torsion rod, (c) upper disc, (d) electrode
fins, (e)lower electrode fin, (f) cell body, (g) base, (h)
annular sample chamber, (i) fill hole, (j) copper foil liner.
The shaded region is machined from a single block of BeCu. The
width of the annular sample space is 1.0 mm but shown exaggerated
for clarity.}\label{schematic}
\end{figure}

Our compound torsional oscillator shown in Fig. \ref{schematic}
was modified from the earlier one\cite{Aoki07,Aoki08a} and has
two torsion rods (a, b) each with diameter 1.9 mm and length 15
mm, and two interconnected masses.  The flange above the upper
rod (a) is rigidly attached to a large copper vibration isolation
block in good thermal contact with the mixing chamber of a
dilution refrigerator.  The upper mass is a "dummy" comprised of
a central disc (c) and two electrode fins (d) made of aluminum
plates.  The TO is capacitively driven by applying dc-biased
sinusoidal voltage between a stationary electrode (not shown) and
one of the movable electrode fins.  The motion of the TO is
detected by measuring the voltage induced between the other
electrode fin biased against another stationary electrode (not
shown).  The lower fin (e) acts as an auxiliary electrode for
measuring the motion of the sample chamber itself.  The lower
mass (f) is mostly made of Stycast 1266 epoxy\cite{Stycast} cast
around the base (g) below the lower torsion rod.  The sample
chamber (h) for solid $^4$He is an annular space (8.0 mm inner
diameter, 10.0 mm outer diameter and 8.0 mm height).  Helium is
introduced into the chamber via the fill hole (i) drilled
(diameter = 0.8 mm) through the center of torsion rods and the
base, and a diametrical channel just below the lower surface of
the base.  To improve the thermal contact between the sample
helium and the mixing chamber, the lower surface of the BeCu base
is pressed against a 100 $\mu$m thick copper foil (j) which is
extended onto the inner wall of the sample space.

The \hesan impurity concentration is increased by adding a
calibrated amount of \hesan gas into the cell prior to loading it
with the commercial ultra high purity \heyon (nominal $x_3$ = 0.3
ppm) at 4.2 K to approximately 80 bar.  A solid sample is
subsequently grown by the blocked capillary method.  The solid
plug formed in the rapidly cooled portion of the filling
capillary maintains a constant mass in the sample space below the
plug.  A Straty-Adams capacitive pressure sensor attached to the
isolation block is used to monitor the pressure in the fill tube
during sample solid formation.  The measured freezing temperature
where a sudden increase in the oscillation amplitude occurs gives
the pressure in the solid formed.  After the plug is formed, the
total time elapsed to freeze \heyon in the sample chamber is
about 40 minutes for all the samples reported here.  Since the
sample solid pressure is 40 bar or greater, where the freezing
temperature is higher than the highest superfluid \heyon
transition temperature, redistribution of \hesan by the heat
flush effect is expected to be unimportant.  Upon completion of
measurements for a given sample, the dilution refrigerator system
is warmed up to near 6 K for pumping out helium from the cell.
The pressure and the TO resonant frequencies are monitored during
this "bake out" procedure to ensure that the remaining residual
amount of helium gas is sufficiently small.  Subsequent to this
procedure, another calibrated amount of \hesan is added such that
the \hesan impurity concentration is larger than the previous
sample and the above procedure is repeated to grow the next solid
sample.  In all, measurements were made on solid \heyon samples
with $x_3$ = 0.3, 3, 6, 12 and 25 ppm.  The samples are
identified by their \hesan impurity concentrations.  The
uncertainty in $x_3$ is estimated to be $\pm$20 \%.

Sample temperature ($T$) is inferred from a ruthenium oxide resistance thermometer attached to the isolation block and is calibrated against a \hesan melting pressure thermometer.  Reproducibility of the ruthenium oxide thermometer upon cycling up to room temperature is verified with a fixed point superconducting standard down to 15 mK.

The two modes of the compound TO may be excited and detected simultaneously.  The resonant frequency and the amplitude of each mode are tracked continuously and independently by two automatic phase-lock feedback data acquisition systems.  In all of the data presented in this report, measurements of both modes are taken simultaneously to ensure identical sample conditions.  The drive levels are set such that sample velocity amplitude of each mode is less than about 15 $\mu$m/s.  It has been found that effects of "critical velocity" and hysteretic behavior are small at these low velocity amplitudes.  The shifts in frequency and amplitudes of the modes can strongly affect each other if the drive levels are increased beyond some threshold velocity.\cite{Keiderling09}

Prior to loading the cell with a sample solid, the "background" characteristics of the TO with no helium in the sample chamber are measured during slow ($\sim$ 15 mK/hr) cooling and verified for reproducibility by following the temperature dependence of the resonant frequencies, $f_{ib}(T)$, and the oscillation amplitudes, $A_{ib}(T)$, of both modes ($i$ = 1, 2).  The background quality factor, $Q_{ib}(T)$, is computed from the tracked oscillation amplitude via $Q_{ib}(T) = (A_{ib}(T)/A_{ib}(T_b))Q_{ib}(T_b)$, where the reference quality factor $Q_{ib}(T_b)$ is determined from measured exponential ring down time (reproducible within 10 \%) at $T_b \approx$ 200 mK.

Background characteristics measured at $T$ = 30 mK are: $f_{1b}$ = 493 Hz, $f_{2b}$ = 1164 Hz, $Q_{1b}$ $\sim$ 7.7$\times$10$^5$ and  $Q_{2b}$ $\sim$ 4.7$\times$10$^5$.  The observed background frequencies are in agreement within $\sim$12 \% with those computed from estimated moments of inertia and torsion constant of the rods.

After a sample of solid \heyon is grown as described above, the resonant frequencies decrease owing to the added moment of inertia.  ``Loading frequency'' $\Delta f_i^0$ in the ``zero'' temperature limit is defined as $\Delta f_i^0 \equiv f_{ib} - f_{is}$ corresponding to the decrease in the mode frequency from the background to sample-filled cell measured near 30 mK.  The measured values are $\Delta f_1^0$ = 0.40 Hz and and $\Delta f_2^0$ = 1.20 Hz for the $x_3$ = 0.3 ppm sample shown in Fig.~\ref{normalized_f1f2}.  These values are within 10 \% of those estimated from the added inertia of solid \heyon in the sample chamber.  The variation in $\Delta f_i^0$ from sample to sample is less than 3 \%.

After a solid \heyon sample is grown in the cell, the same procedure as in the background characterization including the drive levels is followed to measure $f_{is}(T)$, $A_{is}(T)$, $Q_{is}(T_b)$ and $Q_{is}(T)$.  The sample is warmed up at most to 300 mK in the $x_3$ = 0.3 ppm sample, to 1.1 K in the 3, 6 and 12 ppm samples, and to 2.0 K in the 25 ppm sample.  Each sample data set is taken over about 24 hour period while cooling down to $\sim$20 mK from 250 mK in the 0.3 ppm sample and from 1 K in all other samples.  The measured temperature dependence of $f_{is}(T)$ and $A_{is}(T)$ does not vary significantly from one temperature sweep to another if the maximum temperature is within these limits.

\section{results}
\subsection{frequency shift}
\begin{figure}
\includegraphics[width=3in]{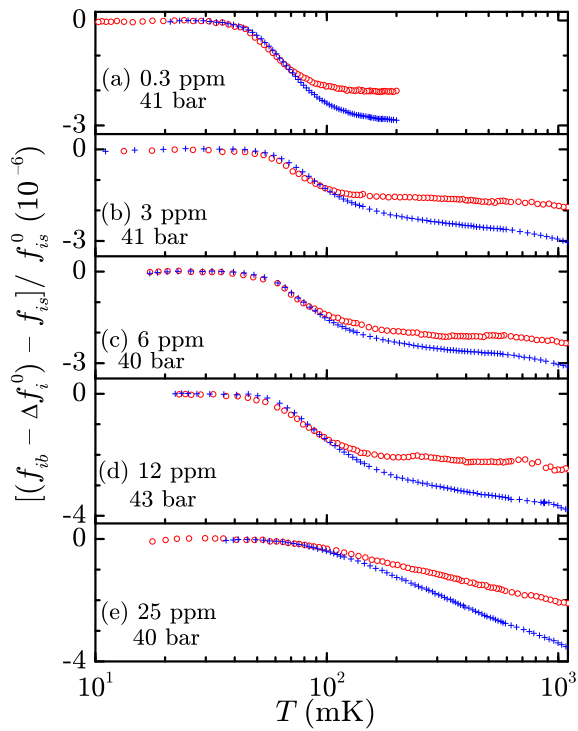}\\
\caption{(color online) Temperature dependence of reduced
frequency shift of the compound torsional oscillator with annular
sample chamber: first mode ((red) open circles) and second mode
((blue) pluses).  Panels (a) $\sim$ (e) show results of solid
\heyon samples containing \hesan impurity concentration of 0.3,
3, 6, 12 and 25 ppm, respectively.  Indicated solid pressures are
estimated from the loading liquid pressure at 4.2 K.}
\label{normalized_f1f2} \end{figure}

Our results on frequency shifts are presented in Fig. \ref{normalized_f1f2} as ``reduced frequency shifts'' defined for
each mode $i$ as:
\begin{equation}
\delta f_i(T)/f_{is}^0 \equiv \frac{[(f_{ib}(T) - \Delta f^0_i) -
f_{is}(T)]}{f_{is}^0},
\label{Eq-reduce-freq}
\end{equation}
where $f^0_{is}$ is the frequency of loaded TO at our minimum
temperature (about 15 mK) depending on $x_3$.  In the zero temperature limit the reduced frequency shift vanishes by definition.  It remains, in all samples, at the zero temperature limit below about 40 mK and its magnitude monotonically increases at higher temperatures for both modes.  Except in the $x_3$ = 0.3 ppm sample, temperature dependence of the reduced frequency shift of the first mode coincides with that of the second mode below about 100 mK.   At temperatures greater than 100 mK changes in reduced frequency of the second mode are greater than those of the first mode.  In all samples, except possibly the 0.3 ppm sample, the reduced frequency shifts for both modes continue to decrease as temperature is increased above 200 mK.  The lack of data in the 0.3 ppm sample prevents us from making a firm statement about the temperature dependence of the reduced frequency shift above 200 mK in this sample.

The temperature range 50 mK $\lesssim T \lesssim$ 150 mK where relatively rapid changes in the reduced frequency shift occurs in Fig.~\ref{normalized_f1f2} has been identified as a signature of the occurrence of NCRI phenomenon.  The reduced frequency shifts would become constant if the observed temperature dependence of $f_{is}(T)$ matched that of $f_{ib}(T)$ at temperatures greater than some "onset" temperature where the fraction of solid sample apparently decoupled from the container, or NCRI fraction (NCRIf), vanishes.  Since identifying such onset temperatures in our samples is ambiguous, reduced frequency shifts rather than NCRIf are shown in Fig. \ref{normalized_f1f2}.

The reduced frequency shift in the 25 ppm sample shows qualitatively distinct behavior from other samples.  In comparison to other samples, there is no temperature range where the reduced frequency shifts vary relatively more rapidly.  At temperatures roughly above 150 mK the frequency shifts decrease linearly (on the logarithmic temperature scale) with the second mode having larger slope than the first.  Measured dissipation in the 25 ppm sample is also distinct from other samples (see Fig. \ref{dissipation}).

\subsection{dissipation}
\begin{figure}
\includegraphics[width=3in]{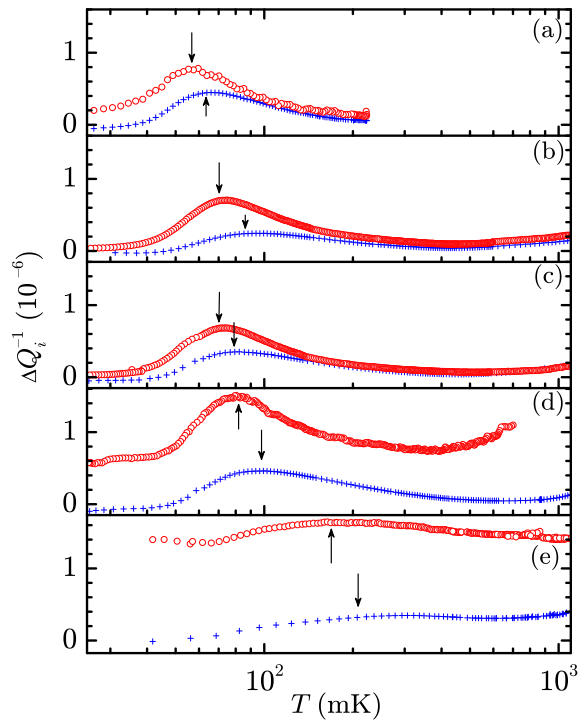}
\caption{(color online) Temperature dependence of the change in
dissipation from empty to loaded sample chamber: first mode
((red) open circles) and second mode ((blue) pluses).
Measurements are taken simultaneously with those reduced
frequency shifts in the samples (with the same panel designations
for \hesan impurity concentration) as shown in Fig.
\ref{normalized_f1f2}. Arrows indicate temperatures where peaks
in dissipation occur.}\label{dissipation}
\end{figure}

The change in dissipation of each mode produced by loading solid \heyon samples is computed by taking the difference:
\begin{equation}
\Delta Q_{i}^{-1}(T) = Q_{is}^{-1}(T) - Q_{ib}^{-1}(T).
\label{Eq-change-dissipation}\end{equation}
Evaluated temperature dependence of $\Delta Q_i^{-1}(T)$ is displayed in Fig.~\ref{dissipation} for each sample shown in Fig.~\ref{normalized_f1f2}.  Samples with $x_3 \leq$ 12 ppm have similar temperature dependence with $\Delta Q_i^{-1}(T)$ passing through local maxima at ``peak temperatures'' ($T_{ip}$) indicated by arrows in Fig. \ref{dissipation}.  The 25 ppm sample shows broader temperature dependence than the other samples. The change in dissipation of the first mode is greater than that of the second in all samples.  This is in contrast to the reduced frequency shift of the second mode being greater than that of the first in all samples at temperatures above 100 mK (see Fig. \ref{normalized_f1f2}).  At temperatures below 40 mK, $\Delta Q_2^{-1}$ for some samples is negative.  This peculiar feature may have resulted in part from an uncertainty ($\sim$10 \%) in the measurements of reference quality factors and from small inaccuracy in the cell temperature measurement below 40 mK.  Small temperature gradients between the sample and the thermometer could lead to errors in taking the difference between the two measurements with the sample chamber being empty and loaded.

\begin{figure}
\includegraphics[width=3in]{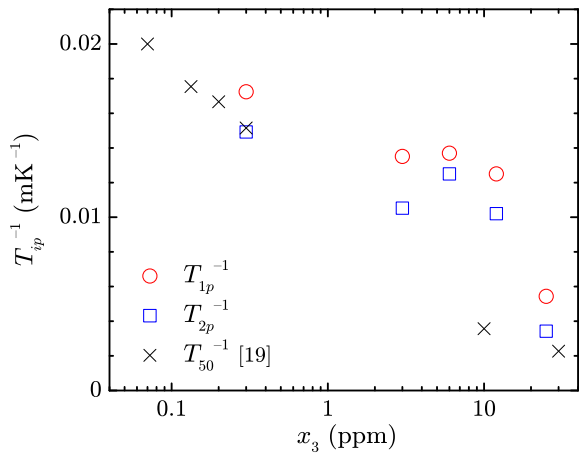}\\
\caption{(color online) Inverse of dissipation peak temperature
vs. \hesan impurity concentration: first mode ((red) circles) and
second mode ((blue) squares).  "Half-maximum temperatures" or
$T_{50}$, (crosses) are taken from Kim, et
al.\cite{Kim08}}\label{T50andTDmax}
\end{figure}

The observed frequency dependent peak temperature that the compound TO technique uniquely yields is an important parameter in considering the dynamical effects occurring in the oscillating \heyon samples.  The peak temperatures are fairly well-defined except in the 6 and 25 ppm samples whose $\Delta Q_{2}^{-1}$ are broader than the others.  The inverse of the peak temperature of the two modes in each sample is shown in Fig.~\ref{T50andTDmax}.  It can be seen that $T_{p2} > T_{p1}$ in all samples.  This frequency dependence of the peak temperature was also found in a cylindrical sample chamber geometry in solid \heyon with nominal 0.3 ppm \hesan concentration\cite{Aoki07} and appears to be independent of sample geometry and size (down to 0.2 mm), \hesan concentration and sample growth conditions.  The peak temperature generally increases for both modes as $x_3$ is increased.  An apparent plateau in $T_{ip}$ in the range 3 ppm $<x_3<$ 10 ppm is likely caused by some accidental variation in as yet unidentified source of dissipation in the sample characteristics.  It is interesting to note that the ``half-maximum temperature'' ($T_{50}$), where Kim, et al.\cite{Kim08} found NCRIf to decrease to half of the maximum at lowest temperatures, smoothly extends the dependence of $T_{ip}$ on $x_3$ found here.

\begin{figure}
\includegraphics[width=3.0in]{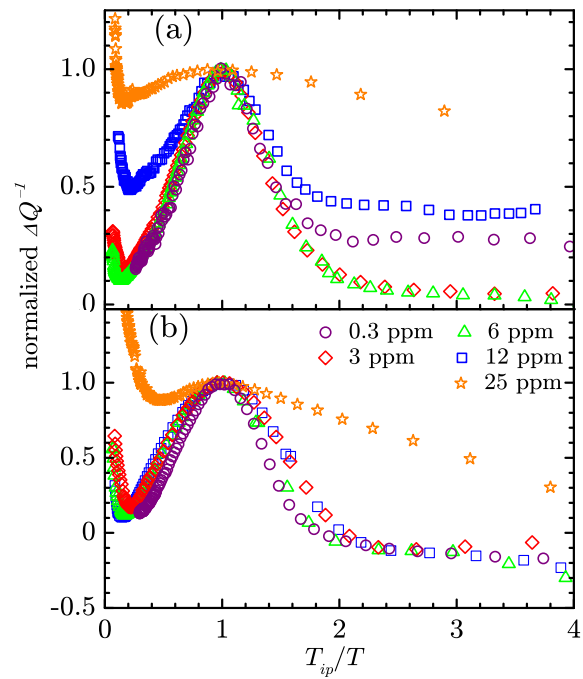}\\
\caption{(color online) Normalized (see text) change in
dissipation: first mode (panel (a)) and second mode (panel (b))
of all samples shown in Fig. \ref{dissipation}.  Symbols:
(purple) circles ($x_3$[ppm] = 0.3), (red) lozenges (3), (green)
triangles (6), (blue) squares (12), and (orange) stars (25).
Results shown for the 25 ppm sample include temperatures up to
2.5 K not displayed in Fig. \ref{dissipation}.}
\label{normalizedDiss}
\end{figure}

The same results as shown in Fig.~\ref{dissipation} are replotted in Fig.~\ref{normalizedDiss} by normalizing the inverse temperature as $T^{-1}/T_{ip}^{-1}$ and the dissipation as $\Delta Q_{i}^{-1}/\Delta Q_{i}(T_{ip})^{-1}$.  Plotting in this manner reveals similarity and dissimilarity among the samples with varying \hesan impurity concentration.  The normalized dissipation for the first mode in the 12 and 25 ppm samples and the second mode in the 25 ppm sample deviate considerably from others.  The "sharp" increases in the normalized dissipation at high temperatures, $T_{ip}/T \lesssim$ 0.2 (first mode) and 0.4 (second mode), become accentuated in Fig. \ref{normalizedDiss}.  Deviations in the samples with higher $x_3$ from the other bell-shaped dependence on $T_{ip}/T$ indicate an emergence of nearly temperature independent ``extra'' dissipation.

Annealing of solid \heyon samples has been found\cite{Rittner06,Penzev07,Reppy10} to affect the frequency shift and dissipation in TO experiments.  Possible effects of sample "annealing" were studied in our $x_3$ = 0.3 ppm sample by raising the temperature of the cell up to 1.8 K (below the melting temperature of the sample) and maintaining the temperature for 10 hours.  The temperature is then reduced to 0.3 K over six hours.  Subsequently measured temperature dependence of the frequency shifts of the two modes are similar to those shown in Fig. \ref{normalized_f1f2} except $f_{is}^0$ decreases by 7.2 and 21.5 mHz for $i$ = 1 and 2, respectively.  This observation is contrary to that of the recent experiment\cite{Reppy10} which finds that the measured frequency shift at lowest temperatures is not affected by annealing.  The dissipation peak temperatures decrease by about 5 mK after annealing.  Our annealing process decreases the extracted (see Sec. \ref{analysis}) characteristic time slightly but produces no significant change in the activation energy.

\section{analysis}\label{analysis}

The original discovery\cite{Kim04a,*Kim04b} of frequency shifts at low temperatures in TOs loaded with solid \heyon and all subsequent confirmations, to our knowledge, have been accompanied by dissipations having puzzling resonance-like temperature dependence as exemplified by panel (a) in Fig. \ref{dissipation}.  In this section, we analyze the observed dependence of dissipations and frequency shifts on \hesan impurity concentration in solid \heyon samples.  It is generally agreed that dissipation peaks occur at temperatures, where an internal dynamical rate ($\tau^{-1}$) matches the imposed TO frequencies ($\omega_i$), or where $\omega_i\tau = 1$.  The internal dynamics of vortex motion\cite{Anderson07,*[see also ]Shevchenko87}, glassy response\cite{Nussinov07}, superglass\cite{Biroli08,Hunt09}, presence of tunneling two level level systems\cite{Andreev09}, and viscoelastic behvior\cite{Yoo09} have been suggested as the physical origin of dissipation.  The vibration of dislocation line segments pinned at network nodes and by \hesan impurity has also been suggested\cite{Iwasa10} as the origin of the observed TO behavior.

\begin{figure}
\includegraphics[width=3.0in]{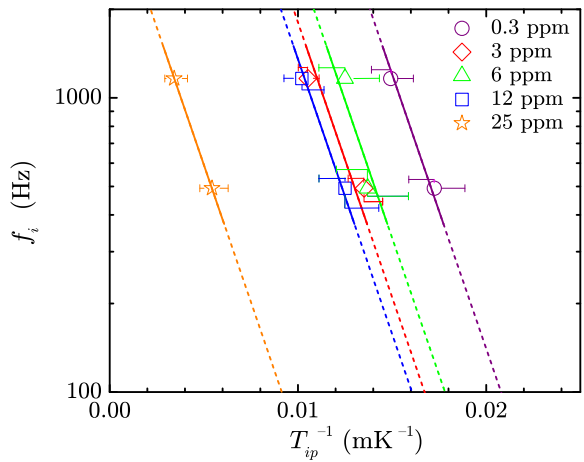}\\
\caption{(color online) Torsional oscillator frequencies plotted
at inverse temperatures where dissipation peaks occur in solid
\heyon samples with given \hesan impurity concentrations.  All
straight lines have the same slope corresponding to an activation
energy $E_0$ of 430 mK and have ordinate intercepts which are
adjusted for best fit for each sample.  The intercepts represent
the characteristic dynamic rate $\tau_0^{-1}$ (see text). Symbols
for samples with different values of $x_3$ are same as those in
Fig. \ref{normalizedDiss}.}\label{Tptau0andE}
\end{figure}

Assuming that the internal dynamics is thermally driven, we consider a simple, activated dynamical time $\tau$ given by the Arrhenius form:
\begin{equation}
\tau = \tau_0\exp \frac{E_0}{k_B (T - T_0)},
\label{Eq-Arrhenius}
\end{equation}
where $k_B$ is the Boltzmann constant, $\tau_0$ a characteristic time and $E_0$ an activation energy.  A possibly non-vanishing ``transition temperature'' $T_0$ is introduced in Eq. (\ref{Eq-Arrhenius}) for generality, but we assume $T_0$ = 0 in our analysis for simplicity.  To examine our dissipation results in terms of Eq. (\ref{Eq-Arrhenius}), the mode frequencies are plotted in Fig. \ref{T50andTDmax} on a logarithmic scale at corresponding $T_{ip}^{-1}$ in each sample.  The slopes and the intercepts determined strictly by straight lines (\emph{not} shown) connecting the two points for each sample in Fig. \ref{Tptau0andE} are ($E_0/k_B$[mK],$\tau_0^{-1}$[s$^{-1}$]) = (380, 5$\times 10^{-7}$), (290, 7$\times 10^{-6}$), (720, 2$\times 10^{-8}$), (370, 3$\times 10^{-6}$) and (430, 3$\times 10^{-5}$) for $x_3$ = 0.3, 3, 6, 12, 25 ppm samples, respectively.  Broadness in the dissipation peaks introduces considerable uncertainties (indicated by error bars in Fig. \ref{Tptau0andE}) in the values of $T_{ip}$ and leads to scattering in these values of $E_0/k_B$ and $\tau_0^{-1}$.
The tendency in Fig. \ref{Tptau0andE}, however, suggests that $E_0$ is a constant independent of $x_3$.  To make progress, let us assume that $E_0/k_B$ = 430 mK, the average of above slopes, is a good estimate for the activation energy.  It is interesting to note that this average activation energy is close to the binding energy of \hesan dislocation lines found in the analysis of their experiments by Kim, et al.\cite{Kim08} and Day and Beamish.\cite{Day07}

\begin{figure}
\includegraphics[width=3in]{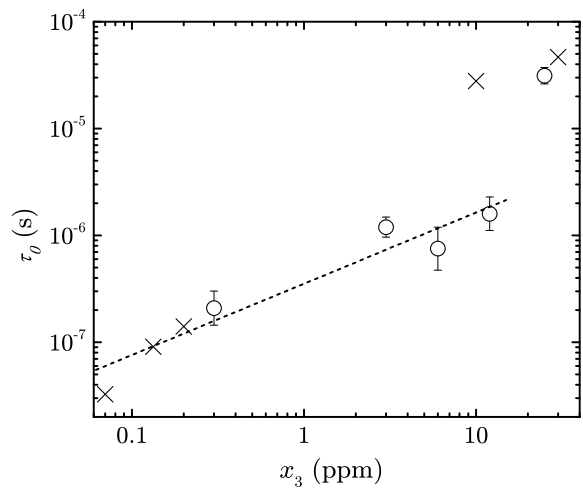}\\
\caption{Characteristic time $\tau_0$ vs \hesan impurity
concentration. The values of $\tau_0$ are determined by the
intercepts of the straight lines in Fig. \ref{Tptau0andE}. The
dashed line represents $\tau_0 = 3.5\times10^{-7}x_3^{2/3}$ s
(see Discussion section of text).  Crosses show $\tau_0$
calculated from the $T_{50}$ data of Kim, et al.\cite{Kim08}
using the same activation energy as in Fig. \ref{Tptau0andE} and
their TO frequencies (see text).}\label{tau0constE}
\end{figure}

The intercepts of the best fits (shown by a straight line for each $x_3$ sample in Fig. \ref{Tptau0andE}) with the set slope specify the values of $\tau_0$ as shown in Fig. \ref{tau0constE}.  There is a trend for $\tau_0$ to increase as $x_3^{\gamma}$ where $\gamma$ $\sim$ 2/3 if $x_3$ $<$ 20 ppm.  Both the reduced frequency shifts and dissipation of 25 ppm sample show qualitatively different behaviors than the samples with lower \hesan impurity concentration. It appears that $\tau_0$ also changes its dependence on $x_3$ beyond 20 $\sim$ 25 ppm.  It was already noted that the values of $T_{50}$ measured by single frequency TO techniques by Kim, et al.\cite{Kim08} are fairly close to $T_{ip}$ (see Fig. \ref{T50andTDmax}).  Constraining straight lines with the same slope as shown in Fig. \ref{Tptau0andE} by their TO frequencies at $T_{50}^{-1}$ imply values of $\tau_0$ as shown in Fig. \ref{tau0constE} for their samples.  Considering differences in sample chamber geometry, sample growth process, measurement methods, etc., $\tau_0$ extracted from $T_{50}$ data of Kim, et al.\cite{Kim08} overlaps and fits surprisingly well with extrapolation of our results to lower values of $x_3$.  Clearly it is of interest to extend frequency dependence studies like ours to smaller \hesan impurity concentration than we have carried out.

\begin{figure}
\includegraphics[width=3in]{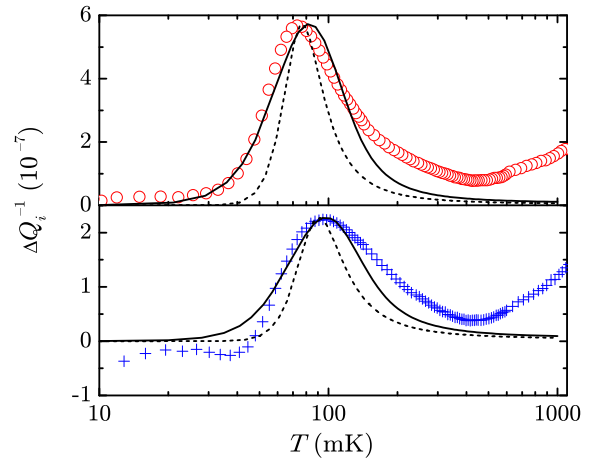}
\caption{(Color online) Comparison of dissipation in the 3 ppm
sample with the dissipation expected from Debye susceptibility.
Same data as shown in Fig. \ref{dissipation} are repeated for
mode 1 ((red) circles) and 2 ((blue) pluses) in panel (a) and
(b), respectively.  In each panel, curves are dissipations
expected (see text) from Eq. (\ref{Eq-DebyeDissipation}):
assuming single activation energy and $\tau_0$ = 1.2$\times
10^{-6}$ s and from Eq. (\ref{Eq-Edistribution}) (dashed curves),
and allowing Gaussian distribution of activation energy with
$w/k_B$ = 120 mK and from Eq. (\ref{Eq-Ddistribtion}) (solid
curves).}\label{dissip_comparison3ppm}
\end{figure}

The mechanical response of driven TOs containing solid \heyon samples has been treated\cite{Nussinov07} by including a general form of rotational susceptibility to account for the sample motion:
\begin{equation}\label{CCsusceptibility}
\chi = \frac{2G}{1 - (j\omega_i\tau)^{\beta}},
\end{equation}
where $G$ is a constant (possibly dependent on frequency\cite{Graf10}), $\omega_i$ is angular frequency, $\tau$ is a relaxation time, and $\beta$ is an exponent dependent on a particular model.  Including the general susceptibility results in extra dissipations and concomitant frequency shifts in the TO response.  In our simplified analysis, the ``Debye model'' response is considered by assuming $\beta = 1$.  In this case, the change in dissipation due to sample motion is given by:
\begin{equation}\label{Eq-DebyeDissipation}
\Delta Q_i^{-1} = \frac{2G\omega_i\tau}{(1 + \omega_i^2\tau^2)},
\end{equation}
and the accompanying reduced frequency shift by:
\begin{equation}\label{Eq-DebyeFreqencyShift}
\frac{f_{is} - f_{ib} - \Delta f_i}{f_{is}^0} = - \frac{G}{(1 +
\omega_i^2\tau^2)}
\end{equation}
Although the linear response model assumed here may not be entirely applicable to our experiment, it is of interest to examine if the model can achieve a similar success as in the description\cite{Syshchenko10} of the shear modulus measurements.
Our results are compared with those expected from Eq. (\ref{Eq-DebyeDissipation}) by assuming the Arrhenius form of relaxation time characterized by our average activation energy and the extracted $\tau_0$ as shown in Fig. \ref{tau0constE}.  Dissipations extracted from the experiment and from the model are compared in Fig. \ref{dissip_comparison3ppm} for the 0.3 ppm sample.  The observed peak dissipation of the first mode at $T_{1p}$ is significantly larger than that of the second mode at $T_{2p}$ in disagreement with the frequency independent peak dissipation expected from Eq. (\ref{Eq-DebyeDissipation}).  To proceed with the (modified) Debye model, the value of $G$ is \emph{adjusted for each mode} separately to match the peak dissipation at $T_{ip}$ in Fig. \ref{dissip_comparison3ppm}.  Dissipations expected from the model produce narrower widths in temperature dependence around the peaks than those observed.  Comparisons in other samples show similar deviations between the observations and the model.  The model of course does not account for the upturns in dissipation at high temperatures nor for the low temperature residual dissipations observed in 12 and 25 ppm samples.

\begin{figure}
\includegraphics[width=3in]{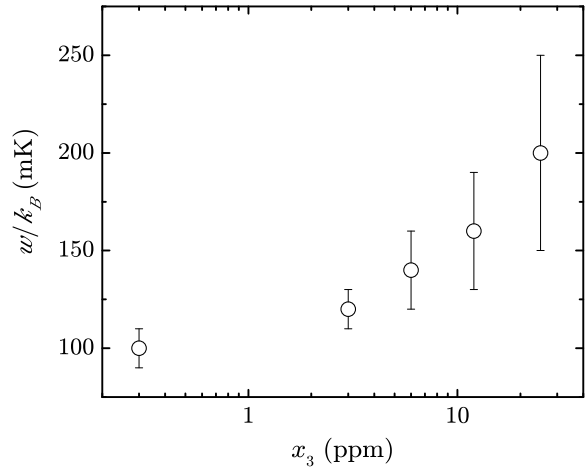}
\caption{Variation of width in distribution of activation energy.
Widths are determined by fitting temperature dependence of
dissipation in each sample using Eq. (\ref{Eq-Edistribution}) and
(\ref{Eq-Ddistribtion}) (see text).}
\label{width}
\end{figure}

As seen in Fig. \ref{dissip_comparison3ppm}, the observed temperature dependence of dissipation is broader than than expected from the Debye model assuming Arrhenius relaxation time with one activation energy for the system.  Broader temperature dependence may be introduced into the model by allowing a distribution in the activation energy.  A canonical Gaussian distribution $N(E)$ given by
\begin{equation}\label{Eq-Edistribution}
N(E) = \frac{1}{\sqrt{2\pi w^2}}\text{e}^{-\frac{1}{2}(\frac{E - E_0}{w})^2}
\end{equation}
is applied to Eq.
(\ref{Eq-DebyeDissipation}) to evaluate $\Delta \bar{Q}_{i}^{-1}$:
\begin{equation}\label{Eq-Ddistribtion}
\Delta \bar{Q}_{i}^{-1}(T) = \int\Delta Q_i^{-1}(T,E) N(E)dE.
\end{equation}
Here, $E_0/k_B$ is set as the average activation energy and the width $w$ of the distribution is adjusted for each sample.  The peak dissipation value $G$ is readjusted \emph{separately} for each mode to match the respective peak dissipation.  For the 3 ppm sample, adjusting to $w/k_B$ = 120 mK can represent the data fairly well as shown in Fig. \ref{dissip_comparison3ppm}.  In the cases of 12 and 25 ppm samples, there appear residual amounts of dissipation at low temperature.  In these two samples, a constant added to $\Delta Q_i^{-1}$ is taken as an additional fitting parameter.   The width of distribution obtained from fitting the data in this manner is shown in Fig. \ref{width}.  Values of $w$ within the error bars shown in Fig. \ref{width} give similar goodness of fit for both modes.  Despite large uncertainties, a clear increasing trend in $w$ is discernable as $x_3$ is increased.  We suggest that  temperature widths around the dissipation peaks in TO experiments can provide sample characterization in the distribution of activation energy.

\begin{figure}
\includegraphics[width=3in]
{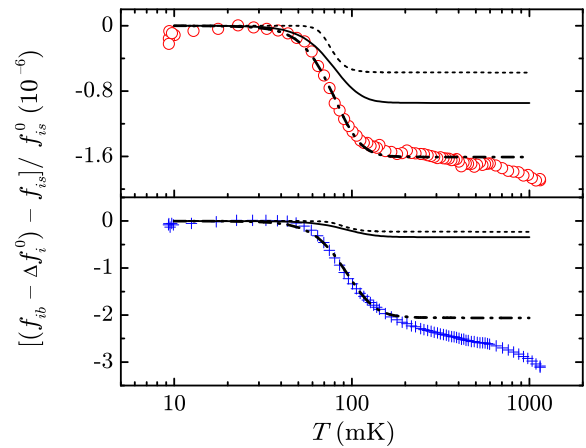}
\caption{(color online) Comparison of $x_3$ = 3 ppm sample
frequency shift data with those expected from Debye
susceptibility using parameters determined from fitting the
dissipation data as shown in Fig. \ref{dissip_comparison3ppm}.
Dashed and solid curves are those expected from the dissipations
shown in Fig. \ref{dissip_comparison3ppm} with the single
activation energy and with the distribution of activation energy,
respectively.  Dash-dotted curves are obtained by multiplying the
solid curves of the respective modes by constants (see text).
Same data (but depopulated for clarity) as shown in Fig.
\ref{normalized_f1f2} are repeated for mode 1 ((red) circles) and
2 ((blue) pluses) in panel (a) and (b),
respectively.}\label{fshift_comparison3ppm}
\end{figure}

Using the same respective parameter values in evaluating dissipation for each mode shown in Fig. \ref{dissip_comparison3ppm}, the frequency shifts expected from Eq. (\ref{Eq-DebyeFreqencyShift}), with single and distributed activation energy, are shown in Fig. \ref{fshift_comparison3ppm} for the 3 ppm sample.  The expected frequency shift of the first mode by including distribution in the activation energy approaches the observed frequency shift but a considerable difference remains.  The expected frequency shifts of the second mode are much smaller than the observed.  The large discrepancy in the second mode occurs despite independently adjusting the value of $G$ for this mode as described above.  The observed decrease in dissipation \emph{and} increase in reduced frequency shift as the TO frequency is increased is a major inconsistency with the modified Debye model above.  The difference in reduced frequency shifts between the measured and the Debye model would be a consequence of a superfluidity in solid \heyonend.   A similar conclusion is made by Yoo and Dorsey\cite{Yoo09} in the analysis of their viscoelastic model for solid \heyonend.

Despite the clear distinction demonstrated in the reduced frequency shifts between the measured and the modified Debye model, similarity in temperature dependence between the two in Fig. \ref{fshift_comparison3ppm} is evident.  Multiplying the Debye expectations shown by solid curves for the first and second mode by constants, 1.7 and 6.0, respectively, gives temperature dependence shown by dash-dotted curves in Fig. \ref{fshift_comparison3ppm}.   Except in high temperature tail regions, the scaled temperature dependence matches the measured reduced frequency shifts quite well.  Similar matching is seen in other samples with multiplicative constants applied to (first mode, second mode) of (1.4, 3.5), (1.3, 3.9), and (1.25, 3.6) in 0.3, 6 and 12 ppm samples, respectively.  The 25 ppm sample cannot be matched in a similar manner owing to the large offset in dissipation.  How this similarity in temperature dependence is related to the interpretation of supersolidity is not yet clear to us.

Nussinov, et al.\cite{Nussinov07} and Graf, et al.\cite{Graf09}
have carried out much more sophisticated analyses of observed TO
responses with generalized rotational susceptibilities expressing
glassy response.  They found that various single frequency TO
responses in both dissipation and frequency shift could be fitted
by appropriately adjusting parameters in the glassy response
model.  Graf, et al.\cite{Graf11} recently reported their study of
the frequency dependence in the dissipation and frequency shifts
observed\cite{Aoki07} in our compound TO with a cylindrical
sample chamber by allowing frequency dependence in the parameter
$G$ in Eq. (\ref{CCsusceptibility}).  In our simple analysis of
the observed \hesan impurity concentration dependent effects in
the present annular sample chamber, we restricted ourselves to
the above modified Debye model in which the value of $G$ is
\emph{adjusted for each mode separately} to account for its
apparent frequency dependence. Comparison of the cylindrical and
annular sample chamber results indicates that the magnitude of
\emph{frequency dependence} of reduced frequency shifts increases
as the sample chamber size is decreased.

\section{Discussion}
%
It is noted that the extracted characteristic time shown in Fig. \ref{tau0constE} varies almost in proportion to $x^{2/3}$ except for the 25 ppm sample.  We speculate that this dependence stems from the diffusion process of \hesan condensed onto dislocation lines in solid \heyon samples.   The diffusion time $\tau_d$ of \hesan along the dislocation lines may be approximated as $\tau_d \approx s^2/D$, where $s$ is some characteristic distance over which \hesan moves during a time interval $\omega_i^{-1}$, and $D\approx lv$ is a diffusion constant with mean free path $l$ and particle velocity $v$.  Let us suppose that the mean free path is approximately given by the effective dislocation loop length $L$.  We follow Iwasa\cite{Iwasa80} in writing $L$ as a parallel combination of the \hesan impurity length $L_i$ and the network pinning length $L_N$:
\begin{eqnarray}
L^{-1} &=& L_i^{-1} + L_N^{-1}\\
&=& \left[g x_3^{-\frac{2}{3}}\exp\left(-\frac{2W_0}{3T}\right)\right]^{-1} +
L_N^{-1},
\end{eqnarray}
where $g = 3.4\times10^{-7}$(see Iwasa\cite{Iwasa10}) is a constant and $W_0$ is the binding energy of \hesan impurity to dislocation line.  At low temperatures where $L_i$ dominates over $L_N$, the characteristic time takes on the Arrhenius form:
\begin{equation}\label{Eq-taufromLi}
\tau_d = \frac{s^2}{g v}x_3^{\frac{2}{3}}\exp\left(E_0/T\right)
\end{equation}
Identifying $\tau_d$ as $\tau$ and the coefficient on the right hand side of Eq. (\ref{Eq-taufromLi}) as $\tau_0$, we expect it to depend on the impurity concentration as $\propto x_3^{2/3}$. Letting  $s^2/v = 1.2\times10^{-9}$ cm s, $W_0/k_B$ ($=3E_0/2k_B$) = 0.65 K, $L_N$ = 2 $\mu$m (similar "fits" can be achieved for 1 $\sim$ 5 $\mu$m) gives the dependence on $x_3$ as shown by the dashed line in Fig. \ref{tau0constE}.  This dependence appears to describe the observed dependence of $\tau_0$ on $x_3$ except for the 25 ppm sample.

There are questions that can be raised on the above diffusion process to describe the observed characteristic time $\tau_0$ as $x_3$ is varied.  If the mean free path of a \hesan atom along a dislocation line is determined by other \hesan atoms condensed onto the dislocation line, it implies that \heyon atoms along the line mysteriously manage not to contribute to scattering.  The process also implies that the diffusion coefficient of \hesan would \emph{decreases} in the low temperature limit as $\exp\left(-E_0/T\right)$.
This temperature dependence is in contrast to the temperature independent quantum diffusion coefficient of \hesan in bulk solid \heyon found by NMR experiments\cite{Allen82} above 0.55 K.  The Kyoto group has observed very long spin-lattice relaxation in their NMR experiment on solid \heyon sample with $x_3 \approx$ 30 ppm.  The long relaxation time might be related to the decreasing $D$ implied by our analysis.

Iwasa\cite{Iwasa10} has analyzed TO experiments and shear modulus shifts in terms of the Granato-L\"ucke model\cite{Granato56a} on the interaction between an externally oscillated TO container and the induced vibrational motion of the dislocation lines present in the loaded \heyon solid sample.  The analysis predicts shifting temperature dependence of the TO frequency and the dissipation peak temperature to higher temperatures as $x_3$ is increased in general agreement with observations.  However, since the expected natural vibration frequencies of the dislocation lines are much higher than the TO frequencies so far attempted, little frequency dependence is expected in both frequency shift and dissipation.  This aspect of the model\cite{Iwasa10} is yet to be reconciled with our TO experiments.

Gaudio, et al.\cite{Gaudio08} proposed a model\cite{Yucesoy10,*Gaudio10} on the effects of \hesan on TO experiments where uniformly distributed \hesan impurities set the maximum grain size in solid \heyon samples.  This model is not likely to be applicable in the range of values of $x_3$ in our experiments where \hesan atoms are expected to condense onto dislocation lines.  Manousakis\cite{Manousakis07} considered \hesan impurity atoms binding to defects and promoting \heyon atoms to interstitial sites.  It is not clear to us how this effect relates to the temperature dependent changes in dissipation observed here as $x_3$ is varied.

Day and Beamish\cite{Day07} discovered that temperature dependent changes in the shear modulus of solid \heyon were almost identical to those in frequency shifts observed in TO experiments.  The discovery gives strong impetus for concluding that the observed changes in shear modulus and frequency shift have similar physical origin.  Syshchenko, et al.\cite{Syshchenko10} recently reported on their measurements of the changes in shear modulus and associated dissipation at frequencies between 0.5 Hz and 8 kHz.  They analyze (see also Su, et al.\cite{Su10a}) their data with a Debye relaxation process and thermally activated dynamics just as we do.  When a distribution of activation energy is included, they find that the relationship between the shear modulus and the dissipation can be accounted for at different frequencies by the assumed Debye and thermal activation processes.  This is in sharp contrast to the analysis of our TO data: including distributions in activation energy cannot account for both frequency shifts and dissipations.  The discrepancy increases at higher frequency (see Fig. \ref{fshift_comparison3ppm}).

\section{conclusion}
Effects of adding \hesan impurity to solid \heyon samples contained in an annular chamber were studied simultaneously at two resonant mode frequencies (differing by a factor of 2.4) by means of a compound torsional oscillator.  Both frequency shifts and extra dissipations produced by the loaded samples were measured.  Maxima in the measured dissipation occurred at impurity-concentration ($x_3$) dependent ``peak temperatures'' around which the frequency shifts varied more rapidly.  When normalized to both the temperature and the dissipation level at the peak, the temperature dependence of dissipation became nearly universal in all samples studied except in the 25 ppm sample.  A thermal activation energy (430 mK) and characteristic relaxation times were extracted from Arrhenius plots of frequency versus the inverse of dissipation peak temperature.  The characteristic time increased with impurity concentration approximately as $ x_3^{2/3}$ and suggested diffusion of \hesan atoms along dislocation lines as the dynamical process producing the observed dissipation.  Observed temperature dependence of dissipation of both modes could be fairly well described by a simple Debye model by allowing for Gaussian distribution of activation energy \emph{if} the magnitude of dissipation at the peak temperature was allowed to be frequency dependent.  The measured magnitudes of frequency shifts were significantly greater than those expected from the model especially in the higher frequency mode.  There remained ``excess'' amounts of frequency shifts which could not be accounted for by the simple Debye model.  The excess frequency shifts may be attributed to superfluidity in solid \heyon at low temperatures.  We believe that these frequency dependent effects hold a key to understanding of the dynamics of quantum solid \heyonend.

\section{acknowledgment}
We thank Moses Chan for stimulating discussions.  This research was supported by NSF DMR-0704120 and DMR-1005325.

\bibliographystyle{apsrev4-1}
\bibliography{supersolid}

\end{document}